\documentclass[letterpaper,12pt]{article}

\usepackage{mathptmx}
\usepackage{authblk}

\usepackage{graphicx}

\usepackage[round]{natbib}%[authoryear,comma,longnamesfirst,sectionbib]

\begin{document}
\newcommand{\pow}{\rm{Pow}}
\newcommand{\pot}{\rm{CP^*}}

\title{An omnibus test for the global null hypothesis}
\author[1]{Andreas Futschik}
\author[2]{Thomas Taus}
\author[3]{Sonja Zehetmayer}

\affil[1]{JKU Linz}
\affil[2]{Vetmeduni, Vienna}
\affil[3]{Medical University of Vienna}
\maketitle

\begin{abstract}
Global hypothesis tests are a useful tool in the context of, e.g, clinical trials, genetic studies or meta analyses, when researchers are not interested in testing individual hypotheses, but in testing whether none of the hypotheses is false. There are several possibilities how to test the global null hypothesis when the individual null hypotheses are independent. If it is assumed that many of the individual null hypotheses are false, combinations tests have been recommended to maximise power. If, however, it is assumed that only one or a few null hypotheses are false, global tests based on individual test statistics are more powerful (e.g., Bonferroni or Simes test). However, usually there is no a-priori knowledge on the number of false individual null hypotheses. We therefore propose an omnibus test based on the combination of p-values. We show that this test yields an impressive overall performance. The proposed method is implemented in the R-package omnibus.  \\

\end{abstract}

\section{Introduction}

When testing multiple hypotheses, the global null hypothesis is often of specific interest. It states that none of the individual null hypotheses is false. In some applications, rejecting the global null can be a goal in itself, whereas in other situation such a test may occur as part of a more sophisticated multiple test procedure. Think for instance of the closure test principle, where the global null needs to be rejected before looking at specific tests. Also, in an ANOVA, the global null is usually tested before testing for pairwise differences.

In meta analysis, rejecting the global null implies an effect at least under some circumstances. Another application is experimental evolution, where
several replicate populations of micro- or higher organisms are maintained under controlled laboratory conditions and their response to selection 
pressures is studied. Further applications where such a test is of interest in its own merit are testing for overall genomic differences in gene expression, and signal detection \citep*[see][]{ING}.

Several approaches to test the global null hypothesis are known. If we assume alternative scenarios where all or most null hypotheses do not hold, combination tests \citep[e.g.,][] {Fisher32,Stouffer}, that sum up two or more independent transformed 
p-values to a single test statistic, have been recommended to maximize power. 
If, however, it is assumed that the null hypothesis holds in most cases, global tests based on individual test statistics are more powerful \citep[e.g., Bonferroni][]{Simes86}. If a larger number of hypotheses is tested, and the alternative hypothesis holds sufficiently often, goodness of fit tests for a uniform distribution of p-values could also be used. They test however for any type of deviation from uniformity, and do not focus specifically on too small p-values. Under more specific models, such as the comparison of several normal means, more specialised tests such as a Tukeys multiple range test, or an ANOVA are further options.

Higher criticism, and checking for {\em overall significance} are alternative terms used instead of global testing. Originating from biblical science, the term {\em higher criticism} was first used by \cite{Tukey} in a statistical context. Making the point that a certain number of falsely rejected null hypotheses can be expected when testing several null hypotheses at level $\alpha$, he then proposed a {\em second level significance test} to check for {overall significance.}
Later Donoho and Jin provided an asymptotic analysis of this and related tests, when the number of hypotheses tends to infinity \citep{Don-Jin1,Don-Jin2} . Their results show that there are situations where there is sufficient power to detect deviations from the global null hypothesis, but no chance to reliably identify in which cases the alternative holds.

Our focus is on a general situation where independent p-values are available from several hypothesis tests that are assumed to be uniformly distributed under the null hypothesis. As there is often no a priori knowledge on the number of false individual null hypotheses, we propose a test that enjoys good power properties, both if few and many null hypotheses are false. Our test is based on cumulative sums of the (possibly transformed) sorted p-values. 

In comparison to other available methods, our simulations show that this test yields an excellent overall behavior. It typically performs better than combination tests, if the alternative holds in only a few cases. If the alternative holds in most cases, it performs better than the Bonferroni and Simes test.
The performance relative to methods that combine evidence across all p-values tends to be even better under those one-sided testing scenarios, where parameters are in the interior of the null hypothesis for some of the tests. 
For these tests, the corresponding p-values will be stochastically larger than uniformly distributed ones, reducing in particular the power of combination tests.

We also present real data applications in the context of meta analysis and experimental evolution.

\section{Testing the global null hypothesis based on p-values}

Consider a multiple testing procedure with $m$ null hypotheses $H_{0i}$, $i=1,\dots,m$. %for the means $\mu_i$, $i=1,\dots,m$ of independent normally distributed observations with known variances $\sigma^2_i$, of which $m_0$ are true and $m_1$ are false. 
We assume that $m$, possibly different, hypotheses tests are carried out leading to stochastically 
independent p-values $p_1,\dots,p_m$. Our focus is on testing the global null \[H_0=\bigcap_{i=1}^m H_{0i},\] 
i.e, that none of the null hypotheses is false. 
We assume that the p-values are either uniformly distributed $$p_i\sim U_{[0,1]}$$ under the global null hypothesis $H_{0}$, or that the p-values are stochastically larger than uniformly distributed ones. In other words, we assume that $P(p_i\le x)\le x$ for $0\le x\le 1.$

Some tests for the global null hypothesis use a combined endpoint, summing up the evidence across all available p-values to a single test statistic (Fisher combination function, Stouffer test). Alternatively other approaches focus on those individual test statistics that lead to extreme p-values, such as in the Bonferroni and Simes tests. As combination tests aggregate evidence across all hypotheses, these tests are particularly powerful when there are (small) effects in many considered null hypotheses. When there are only a few (large) effects, global tests based on individual test statistics are more powerful. Other approaches are goodness of fit tests or higher criticism.

\subsection{Omnibus test}

%Let $Z_i=\bar{X}_i/\sigma_i \sqrt n$ denote individual z-test statistics for hypotheses $H_{0i}$, $i=1,\dots,m$, where $\bar{X}_i$ denotes the arithmetic mean of the observations for hypothesis $i$ and $n$ the corresponding sample size. 
\paragraph{General outline}
Starting with independent p-values $p_1,\dots,p_m$, we
denote the sorted p-values by $$p_{(1)} \leq \dots \leq p_{(m)},$$ and transform them with a monotonously decreasing function $h(\cdot)$ so that small 
p-values lead to large scores. 
Possible choices for $h(\cdot)$ will be discussed below.
Next we obtain the L-statistics $S_i=\sum_{j=1}^i h(p_{(j)})$, $i=1,\dots,m$. 
Each of these partial sums could in principle be chosen as a test statistic for the global test and the 
best choice in terms of power for a specific scenario will depend both on $(m_0,m_1)$ and the respective effect sizes.
Since these quantities are unknown, 
we propose to select the most unusual test statistic out of $S_i$, $1\le i\le m$.
If the scores $S_i$ were approximately normally distributed, we could standardize them to figure out how unusual they are.
Here however, the distribution of the $S_i$ with small index will be closer to an extreme value distribution, Therefore, 
we transform the sums using the distribution function $G_i$ of $S_i$ 
under the global null hypothesis, and take
\[ 
T^{\ast}=\max_{1\le i\le m} G_i(S_i).
 \]
as our test statistic. 
Although the cumulative sums $S_i$ may be viewed as L-statistic, and 
conditions that ensure the asymptotic normality of L-statistics $m\rightarrow\infty$ are known (see e.g.\cite{Stigler69},
these conditions are not satisfied for some of the $S_i$, and furthermore
the number of hypotheses is small to moderate.
We therefore estimate the distribution of $T^\ast$ by simulating uniformly distributed p-values under the global null.
Notice however that for some underlying distributions of $h(p_j)$, 
such as uniform, exponential or (skewed) normal, 
exact distributions are available for $S_i$ (\citep{Crocetta}, \citep{Nagaraja2006} ). 

Later on, we will consider four transformations $h(p)$ in more detail:
\begin{itemize}
\item $h(p)=1-p$ (omnibus p)
\item $h(p)=-\log p$ (omnibus log p) 
\item $h(p)=\Phi^{-1} (1-p)$ (omnibus z)
\item $h(p)=p^{-\alpha}$ with $\alpha=0.5$ (omnibus power). 
\end{itemize}

Notice that for small enough $p$, we have that
\[ 1-p
\le \Phi^{-1} (1-p)\approx \sqrt{2\log(1/p)}
\le \log (1/p)\le p^{-\alpha}.
  \]
Thus different choices of $h(.)$ assign different relative weights to small p-values.

\subsection{Alternative test statistics}

We briefly explain the most popular approaches that use p-values for testing the global null hypothesis.

\paragraph{Fisher combination test} 

\citet{Fisher32} proposed the combined test statistic given by $T=-\sum_{i=1}^m 2 \log p_i$. 
Under the assumption of independent uniformly distributed p-values, 
the null distribution is $T \sim \chi^2_{2m}$.  

\paragraph{Stouffer's z} 

Based on z-values $Z_i=z_{1-p_i}$, where $z_{1-p_i}$ denotes the $1-p_i$ quantile of the standard normal distribution, the combined test statistic is given by $Z=\sum_{i=1}^m Z_i/\sqrt m$. Assuming again independent uniformly distributed p-values under the global null, 
it can be easily seen that $Z \sim N(0,1)$ \citep{Stouffer}. 
%An attractive feature of Stouffer's combination test is that weights for the individual hypothesis tests are easy to introduce. 
%With weights, the combined test 
%becomes $Z_w=\sum_{i=1}^k w_i Z_i/\sqrt {\sum_{i=1}^k w_i^2}$.

\paragraph{Bonferroni test} 

The Bonferroni test rejects the global null hypothesis, if the minimum p-value falls below $\alpha/m$, 
i.e.\  $\min_i p_i \leq \alpha/m$ \citep[see, e.g.,][]{Dickhaus}. The Bonferroni test controls the family-wise error rate at level $\alpha$ in the strong sense. The test 
makes no assumption on the dependence structure of endpoints. For independent test statistics, $\alpha/m$ may be replaced by the slightly more 
liberal upper bound $1-(1-\alpha)^{1/m}.$

\paragraph{Simes test} 

An improvement of the Bonferroni test in terms of power was proposed by \cite{Simes86}. For the $m$ hypotheses $H_{0i}$, $i=1,\dots,m$ with p-values $p_i$, the Simes test rejects the global null hypothesis if for some $k=1,\dots,m$, $p_{(k)} \leq \alpha k /m$. In the last decades the Simes test has become very popular for testing individual hypotheses controlling the False Discovery Rate.

\paragraph{Higher criticism}

Based on an idea by \cite{Tukey}, \cite{Don-Jin1,Don-Jin2} introduced the higher criticism HC to test the global null hypothesis of no effect for independent hypotheses. It is defined by $$HC^*_m=\max_{1 \leq i \leq \alpha_0m} \left\{ \sqrt m \frac{i/m-p_{(i)}}{\sqrt{p_{(i)}(1-p_{(i)})}} \right\}.$$ $\alpha_0$ is a tuning parameter often set to 1/2, and has been studied in 
particular for large scale testing problems.

\paragraph{Goodness of fit tests} 

For our global test problem of independent p-values and under a point null hypothesis,
the p-values $p_i$, $i=1,\dots,m$, usually follow a uniform distribution $U(0,1)$. Thus any goodness 
of fit test for uniformity, such as the Kolmogorov-Smirnov (KS), the 
Chi-square and the Cramer-von Mises tests also provide tests for the global null hypothesis. 
The KS test, for example, would use the maximum distance between the empirical distribution function of the observed p-values and the uniform
distribution function, $D_n=sup_{0\le x\le 1} |F_n(x)-x|$ as test statistic. A disadvantage of goodness of fit tests in our context is
 that they test not only for smaller than expected p-values but against any deviation from uniformity. As also confirmed by our simulations, these 
tests therefore provide lower power compared to more specialized tests in our situation (data not shown).

\section{Results}

%\subsection{Setup of the simulation study}
We start our simulation study by comparing the power of our test when different transformations $h(\cdot)$ are used.
It will turn out that $h(p)=-\log p$ leads to a particularly good overall behavior across different scenarios, 
and we thus focus on this transformation when comparing our approach with alternative tests for the global null, such as the 
Bonferroni and the Simes procedure, as well as Fisher's and Stouffer's combination test. 
Although typically used for a large number of hypotheses, we will also consider higher criticism as a competing method (with the tuning parameter $\alpha_0=0.5$). As the asymptotic approximations do not necessarily hold for small numbers of hypotheses, we simulate critical values under the null model for this test.

We simulate different scenarios by varying both the total number $m$ of hypotheses, and
the number $m_1$ of instances where the alternative holds. 
We assume independence between the p-values, which was a condition in our derivation of the omnibus test.

Although our test is based on p-values that may arise in a multitude of settings, we 
want to specify effect sizes and alternative distributions in an intuitive way,
and therefore compute our p-values from normally distributed data with known variance $\sigma^2=1$ and equal sample sizes $n$. 
More specifically, we consider the one-sample z-test for one-sided hypotheses 
$$H_{0i}: \mu_i=0 \quad \mbox{versus}\quad H_{1i}:\, \mu_i>0, \; i=1, \dots, m,$$ 
for the mean of the observations. 

In the simulations, we first assume that all alternatives have the same mean effect $\Delta/\sigma$ and for the true null hypotheses $\Delta=0$. 
Later on we also consider the following setups: 
\begin{description}
\item[(i)] Negative effect sizes that are in the interior of the null hypotheses: We assume that under the true null hypothesis, 
the data have a negative effect size of $-\Delta/\sigma$ and under the alternative hypotheses a positive effect size of $\Delta/\sigma$. 
\item[(ii)] Different effect sizes of alternative hypotheses: 
We assume randomly chosen exponentially distributed effect sizes with a rate parameter of $3 \sqrt m_1$. 
\item[(iii)] Different effect sizes of alternative hypotheses and different effect sizes in the interior of the null hypotheses: 
We assume randomly chosen exponentially distributed effect sizes with a rate parameter of $3 \sqrt m_1$ or $-3 \sqrt m_1$, respectively.
\end{description}

All computations were performed using the statistical language R \citep{R}, the Fisher and the Stouffer combination test were calculated using the 
function combine.test in the survcomp package \citep{survcomp}.
For each scenario at least 10000 simulation runs were performed.

For all following simulation results, the methods control the Type I error of $5\%$ if the global null hypothesis is true (data not shown). 

%***ueberpruefen!!***

\subsection{Influence of the chosen transformation on the omnibus method}

Fig. 1 shows power curves for the omnibus test using the four proposed transformations. We consider $m =10$, $m_1 \in \{1,3,5,10\}$, and $\Delta/\sigma=0.3/\sqrt m_1$. These variants show similar power values for a lot of scenarios. 
Nevertheless the performance of the power (``power'') and identity transforms (``p'') seems to be somewhat less satisfactory. In particular the power transform 
performs considerably worse when the alternative is true in several instances, while giving only slightly better results in 
the case of only one true alternative. The 
{\em z} and {\em log p} transforms both show a good overall behavior. The {\em log p} transform performs
slightly better for small $m_1$ (i.e. a few larger effects), whereas
the omnibus {\em z} method turns out to be slightly better if $m_1$ is large (i.e. several smaller effects). 
In section \ref{powcomp} we provide a between methods comparison 
of the worst case power across all possible choices of $m_1$ with constant cumulative effect sizes.
According to Table~\ref{tab1}, the omnibus {\em log p} transform slightly outperforms the $z$ transform. Thus
we will use the {\em log p} transformation with our omnibus test subsequently.

\begin{figure}[!ht]%figure1
\centering
\includegraphics[width=0.9\textwidth]{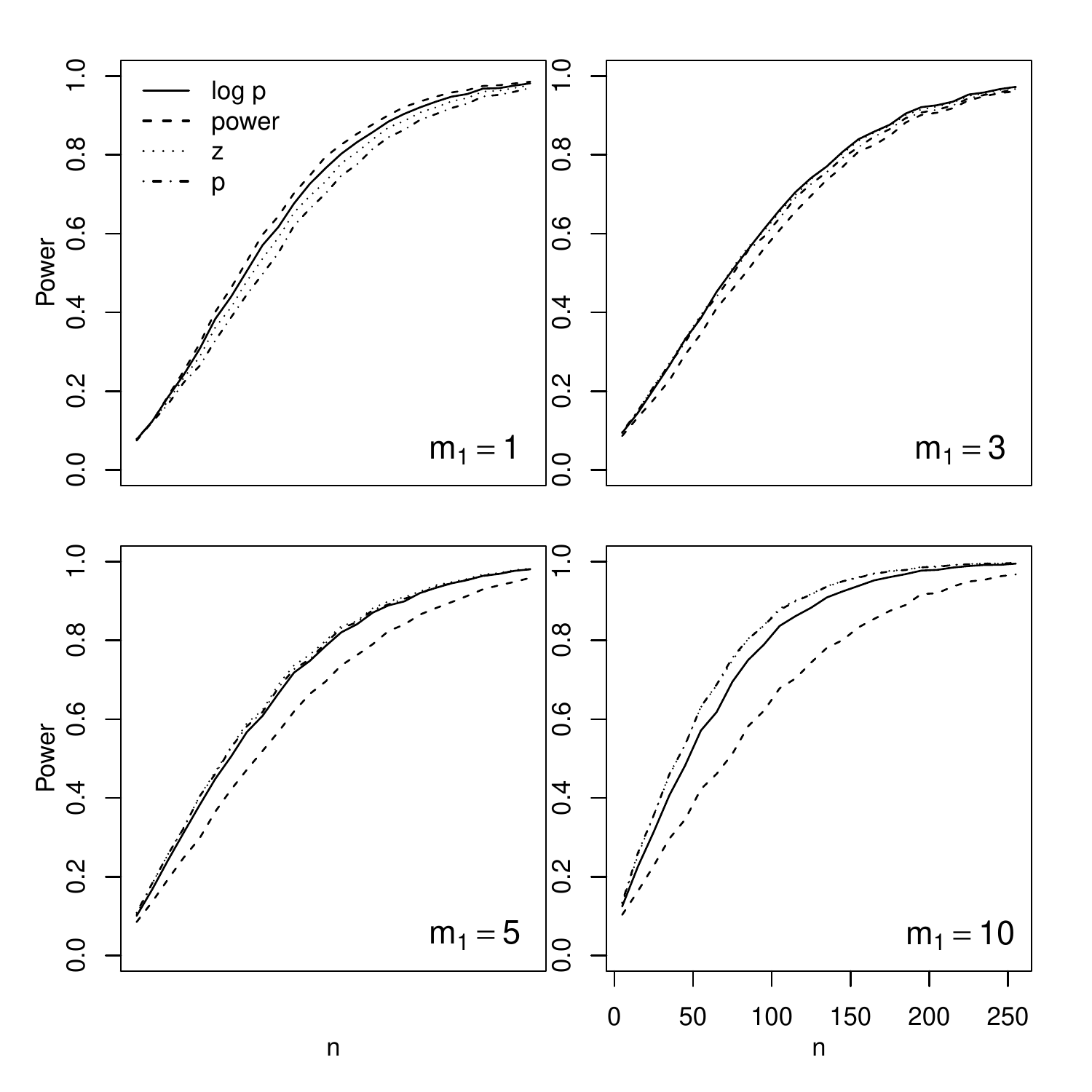}% [width=16.5cm,height=16cm]
\caption{Power values for omnibus log p, power, z, and p are given for increasing $n$, $m=10$, $m_1 \in \{1,3,5,10\}$, $\Delta/\sigma=0.3/\sqrt m_1$.}\label{fig:01}
\end{figure}

\subsection{Power comparison between different testing methods \label{powcomp} }

\begin{figure}[!ht]%figure1
\centering
\includegraphics[width=0.9\textwidth]{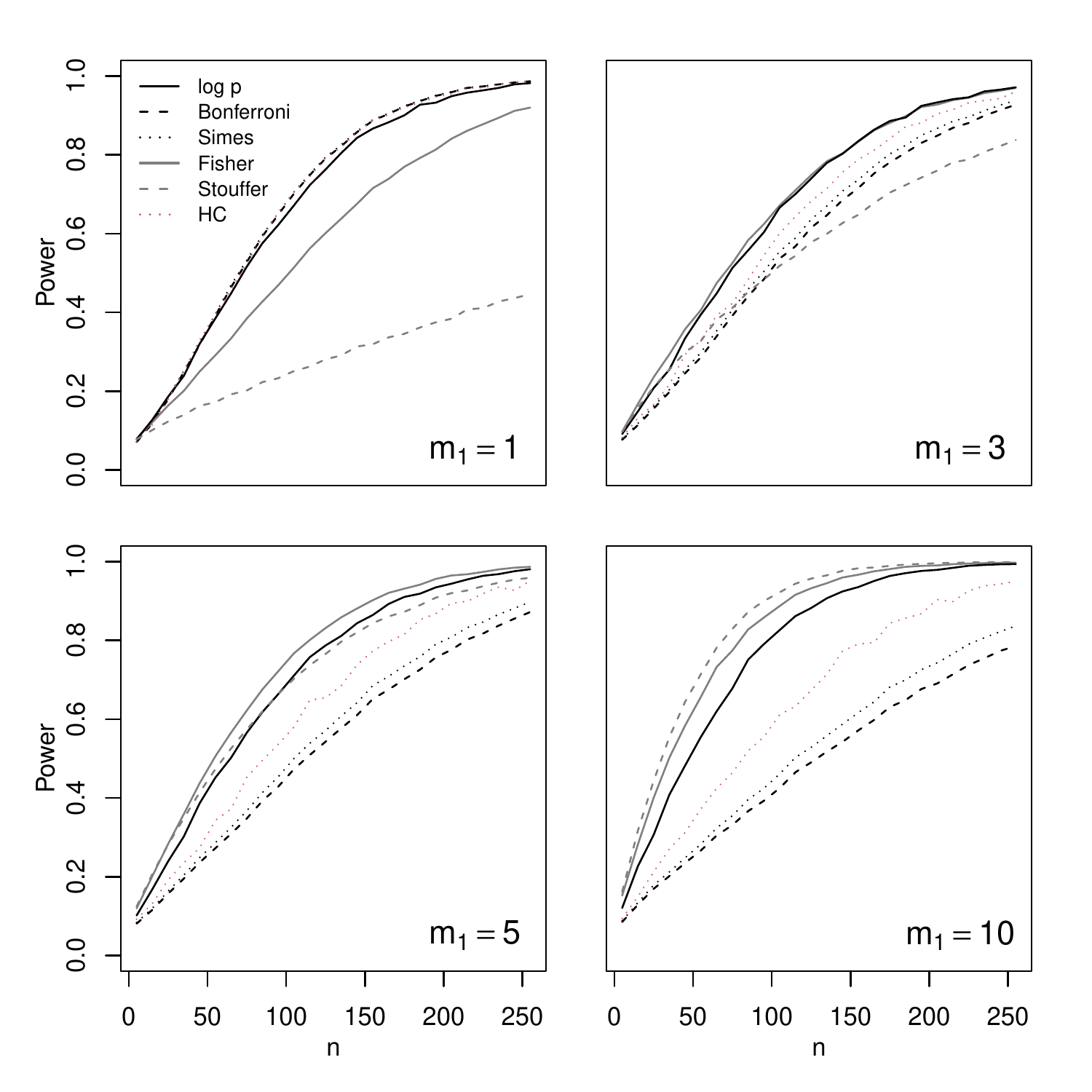}% [width=10.5cm,height=16cm]%[width=0.8\textwidth]
\caption{Power values for increasing $n$, $m=10$, $m_1 \in \{1,3,5,10\}$, $\Delta/\sigma=0.3 / \sqrt m_1$ for omnibus log p, Bonferroni, Simes, Fisher, and Stouffer test, HC test)}\label{fig:02}
\end{figure}

Figure \ref{fig:02} shows power curves for omnibus log p, Bonferroni, Simes, Fisher, Stouffer test, and HC for $m =10$, $m_1 \in \{1,3,5,10\}$, and $\Delta/\sigma=0.3/ \sqrt m_1$. It can be seen that the omnibus method is among the top methods concerning power for all scenarios (black solid curves). 

The Bonferroni and Simes methods give the best power results in the case of only one false null hypothesis, $m_1=1$, however, the difference to the omnibus log p variant of our test is only marginal. 
For increasing $m_1$ the power of the Bonferroni and Simes methods is inferior compared to all other methods. 
As expected the Simes test outperforms the Bonferroni procedure (or is equal), though, for the considered scenarios the improvement in power is only small. 

The Fisher combination test is slightly superior in scenarios with large $m_1$ in comparison to the omnibus tests, however, it has low power for small $m_1$. E.g, for scenarios with $m_1=1$ the omnibus test has nearly 20 percentage points higher power than the Fisher test. The Stouffer test only shows competitive power values for high number of false null hypotheses for the considered scenarios. In contrast the HC method for $\alpha_0=0.5$ has similar power values as Bonferroni and Simes for $m_1=1$, for increasing $m_1$ the omnibus log p, Fisher, and Stouffer test are clearly more powerful.

\paragraph{Worst case behavior}

We assess also the overall behavior of the statistical tests we considered by looking at the minimax power across scenarios that involve all possible numbers $m_1$
of true alternative hypothesis. We define the minimax power as then the lowest power across all these scenarios.
With $m_1$ alternatives, the individual effect size was chosen $\Delta/\sigma=\gamma/\sqrt m_1$. 
This leads to a constant cumulative effect size of 
$\frac{\sqrt{m_1}\gamma}{\sqrt{m_1\sigma^2}}=\gamma/\sigma^2$. This constant cumulative effect size would lead to equal power for any value of $m_1$ with a likelihood ratio test in the simplified scenario assuming $m_1$ null hypotheses that are either all true or false. In the theoretical case that $m_1$ and the position of the $m_1$ hypotheses is known, 
an optimal test could be obtained this way that leads to constant non-centrality parameters for all values of $m_1$.
The below table uses $\gamma=0.3$, leading to intermediate power values. As can be seen, the considered omnibus tests outperform the other 
tests with respect to the worst case behavior, with the omnibus {\em log p} test performing best.

 \begin{table}[h!]
\caption{\label{tab1}
Minimax power. Worst case power values for $m_1$ from 1 to $m$ (minimum over all simulation scenarios) for $n=\{100,200\}$, $m=\{10,20,1000\}$, $\Delta/\sigma=0.3/\sqrt m_1$.
}
\begin{tabular}{c|cccccccccc}
&\multicolumn{2}{c}{$m=10$} & \multicolumn{2}{c}{$m=20$} & \multicolumn{2}{c}{$m=1000$}\\
& $n=100$&$n=200$& $n=100$&$n=200$& $n=100$&$n=200$\\ \hline
Omnibus log p & 0.63& 0.92& 0.50& 0.84& 0.23& 0.50  \\
Omnibus z     & 0.62& 0.92& 0.49& 0.83& 0.23& 0.49   \\
Omnibus p     & 0.59& 0.90& 0.46& 0.82& 0.22& 0.48 	\\
Bonferroni    & 0.41& 0.68& 0.28& 0.47& 0.11& 0.18    \\
Simes         & 0.44& 0.73& 0.30& 0.52& 0.12& 0.19       \\
Fisher        & 0.49& 0.83& 0.35& 0.67& 0.15& 0.28        \\
Stouffer      & 0.24& 0.39& 0.16& 0.25& 0.09& 0.11  \\
HC half       & 0.53& 0.86& 0.40& 0.73& 0.14& 0.30  \\
\end{tabular}
\end{table}

\paragraph{Behavior for small numbers $m_1$ of true alternatives}

To further compare the power of our omnibus {\em log p} test and Fisher's test, we performed simulations when $m_1$ is small, 
either in absolute terms or compared to $m$. More specifically we considered $m_1=1$, $m_1=5$, as well as $m_1=m/10.$
We assigned the same fixed effect sizes $\Delta/\sigma \in \{0.25, 0.5\}$ to each alternative hypothesis.
Figure 3 shows the power curves of the omnibus test (black curves) and the Fisher combination test (grey curves) 
for $n \in \{20,40\}$, and increasing $m$. The omnibus test provides a higher power in most scenarios. 
Only in the situation of small effect sizes ($\Delta=0.25$), the Fisher combination test
behaves better under some circumstances. This occurs in particular when $m_1=5$, and $m$ fairly small, implying a fairly large 
proportion $m_1/m$ of alternatives.  
Notice however, that the difference in power is small in these cases compared to the excess power of the omnibus test for larger effect sizes.

\begin{figure}[!ht]%figure2
\centering
\includegraphics[width=0.9\textwidth]{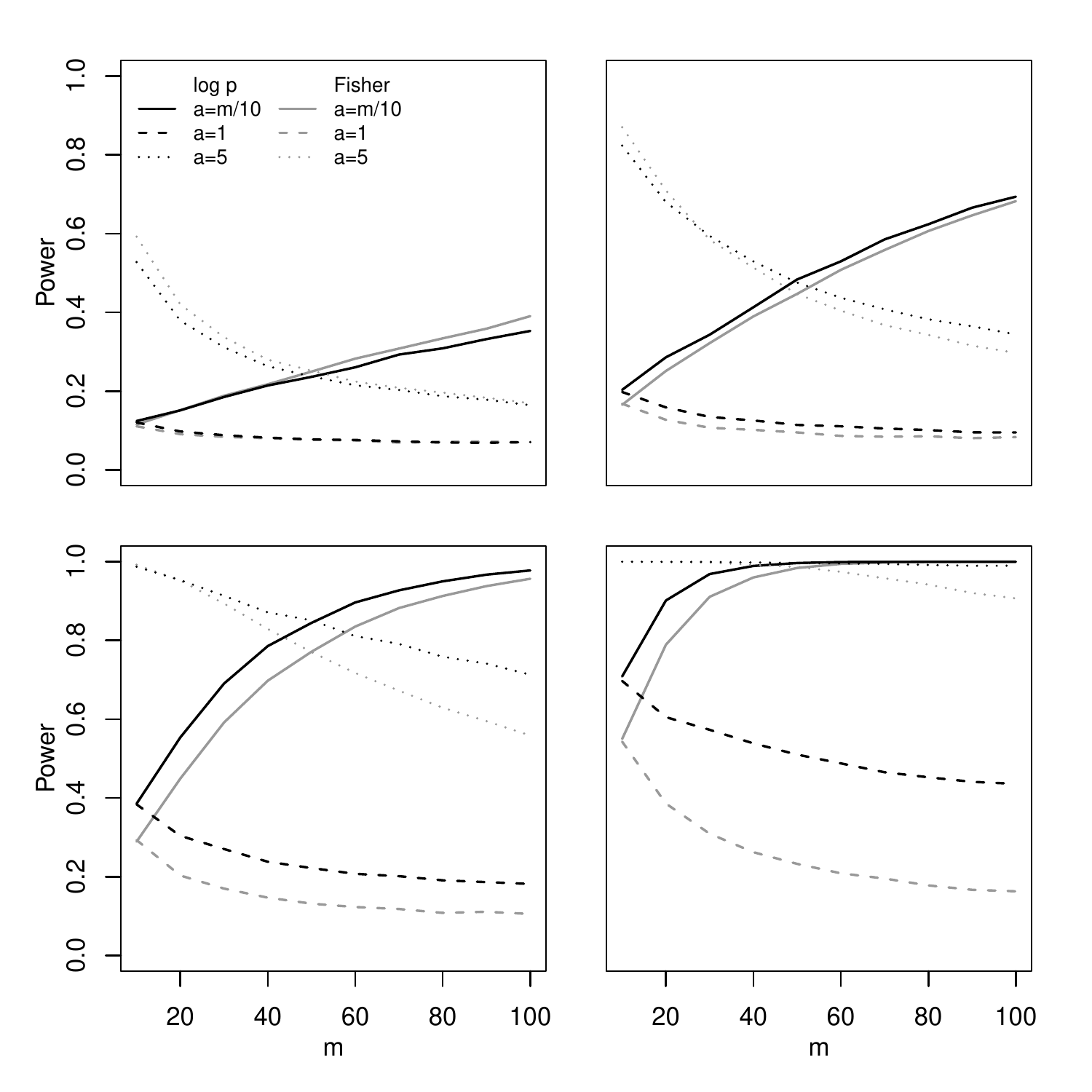}% [width=10.5cm,height=16cm]%[width=0.8\textwidth]
\caption{Omnibus log p (black line) and the Fisher combination test (grey line) for $n \in \{20,40\}$, $\Delta \in \{0.25, 0.5\}$, increasing $m$ and $m_1=m/10$, $m_1=1$, or $m_1=5$, respectively. }\label{fig:03}
\end{figure}

\subsection{Distributed/negative effect sizes}

In Fig. 4 (first row) we show simulation results for distributed effect sizes with a mean effect $\Delta$ distributed according to an exponential distribution with a rate parameter of $3 \sqrt m_1 $ for $m_1 \in \{1,3,5\}$, $m=10$. Generally, the power values are much lower than for equal mean effect sizes. Still, the omnibus log p method has maximum power in nearly all scenarios, only for $m_1=10$ the Fisher combination test is more powerful.

\begin{figure}[!ht]%figure1
\centering
\includegraphics[width=1\textwidth]{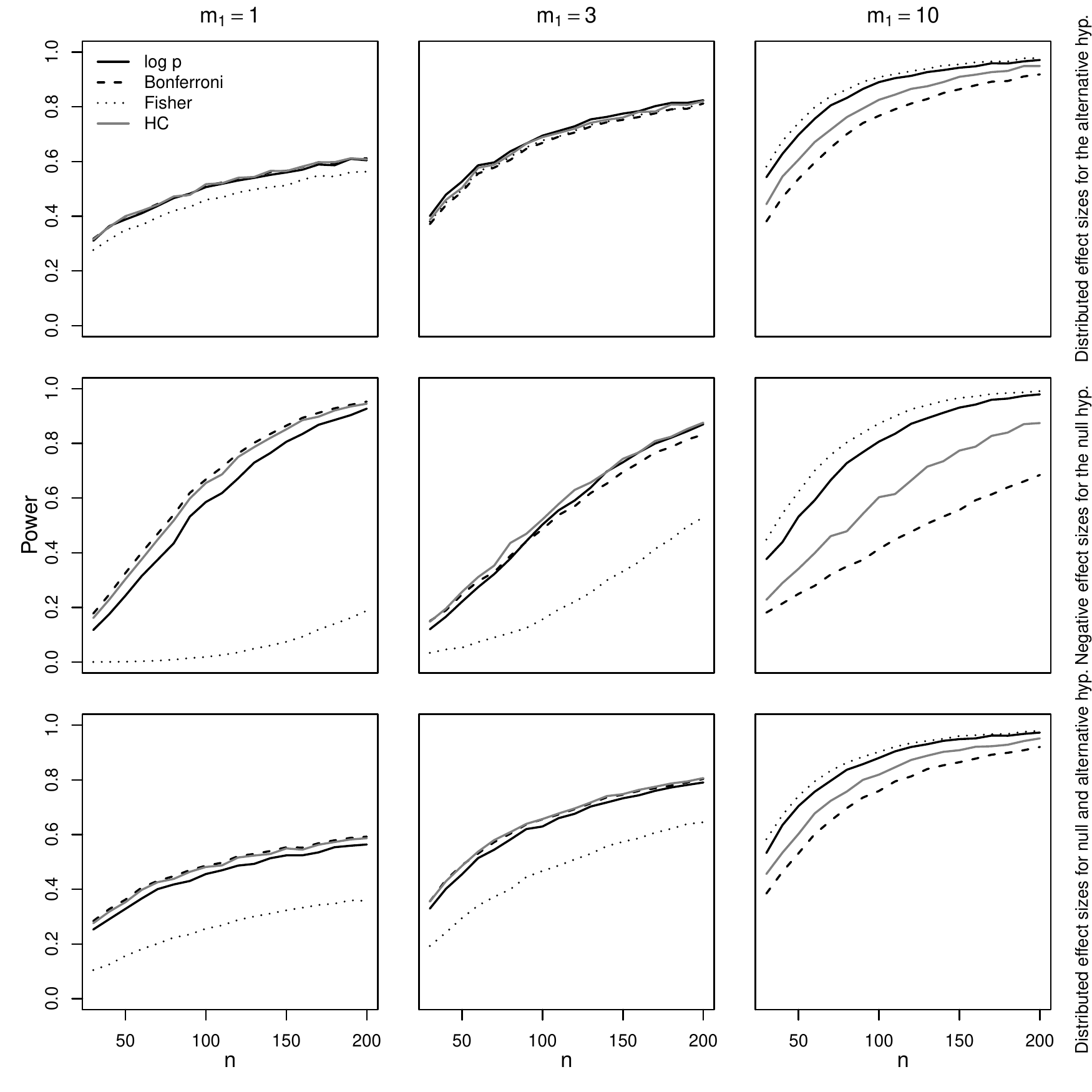}% [width=10.5cm,height=16cm]%[width=0.8\textwidth]
\caption{
Power values are given for increasing $n$, $m=10$, $m_1 \in \{1,3,10\}$ for omnibus log p, Bonferroni, Fisher, and HC. The first row shows results for distributed effect sizes of alternative hypotheses according to an exponential distribution with rate parameter $3 \sqrt m_1 $. The second row shows results for
$\Delta/\sigma=-0.3/ \sqrt m_1$ under the null hypothesis and $\Delta/\sigma = 0.3/ \sqrt m_1$ under the alternative. 
The third row shows results for distributed effect sizes according to an exponential distribution under the alternative as well as under the null hypothesis. }\label{fig:04}
\end{figure}

Fig. 4 (second row) shows results for negative effect sizes under the null hypothesis, leading to p-values that are
stochastically larger than uniform. A comparison with Figure 2 reveals that this does not much
influence the power of the omnibus test ($\Delta=0.3 \sqrt m_1$ for $m_1 \in \{1,3,5\}$, $m=10$), but it reduces the power of the Fisher combination
test a lot for small $m_1$. The same is true for the Stouffer test (not shown), as it also uses the sum over all (transformed) p-values. 
The power difference between the omnibus test and the Fisher combination test reaches more than 70 percentage points for, e.g., $m=10$, $m_1=1$, $\Delta=0.3$. %We also found scenarios where the power of the omnibus test was 1 compared to a power of 0 for the Fisher combination test (e.g., $m=100$, $n=40$, $\Delta=0.5$, and $a=10$ (data not shown). 
The power of the Bonferroni test and of higher criticism changes even less compared to the omnibus test
when parameters are in the interior of the null hypothesis. 

If both alternative and null hypotheses have effect sizes distributed according to an exponential distribution (with a rate parameter of $3\sqrt m_1$ 
for alternative hypotheses and $-3\sqrt m_1$ for null hypotheses), the relative behavior of the methods (Fig. 4, third row) is qualitatively similar 
to that implied in the second row. As observed in the first row however, the power clearly decreases for all methods with randomly distributed 
effect sizes.

\subsection{P-values from discrete data}

The assumption of uniformly distributed p-values 
under the null hypothesis is not always satisfied. Besides the possibility of parameter values
in the interior of the null hypothesis, also discrete models lead to p-values that are not uniformly distributed on the interval $[0,1]$.
To also cover the case of discrete data, we performed a simulation study under a two sample binomial model. 
For the first group, the simulated data were $B(n,p_0)$ distributed, for the second group, again generated from $B(n,p_0)$ under the
null hypothesis and from $B(n,p_1)$ under the alternative. Here, $n$ denotes the per-group sample size. A $\chi^2$-test with one degree of freedom was performed and a corresponding p-value was calculated. If both groups showed only successes or only failures, the p-value was set to $p=1$. 

We first checked whether the type I error is still controlled under our discrete model. For this purpose we considered sample sizes $n$ between 10 
and 100, as well as allele frequencies $p_0$ in $[0.05,\,0.5]$ under the null hypothesis. 
Although for small $n$ the distribution of the test statistic is not well approximated by the 
chi-square distribution, we nevertheless used the standard p-values produced by the R function {\em chisq.test}. 
Our simulations showed no violations of the type I error probability of $\alpha=0.05$. This is since the chi-square test tends to become 
conservative (and the p-values stochastically larger than uniform) for small $n$. In other testing situations where this is not the case, 
type error control may however be an issue.

Figure 5 provides the power obtained when using our omnibus test on several scenarios.
The left plot shows the power values as a function of $n$ from 10 to 100 for omnibus log p for $m=10$, $m_1\in \{1,3,5,10\}$. The plot in the middle shows the power values as a function of $p_0$ with constant $n=50$ and $p_1=p_0+0.2$ increasing in the same amount as $p_0$. For the right plot, $p_0=0.4$, $n=50$ and $p_1$ is increasing from 0.4 to 0.9.

\begin{figure}[!ht]%figure2
\centering
\includegraphics[width=1.0\textwidth]{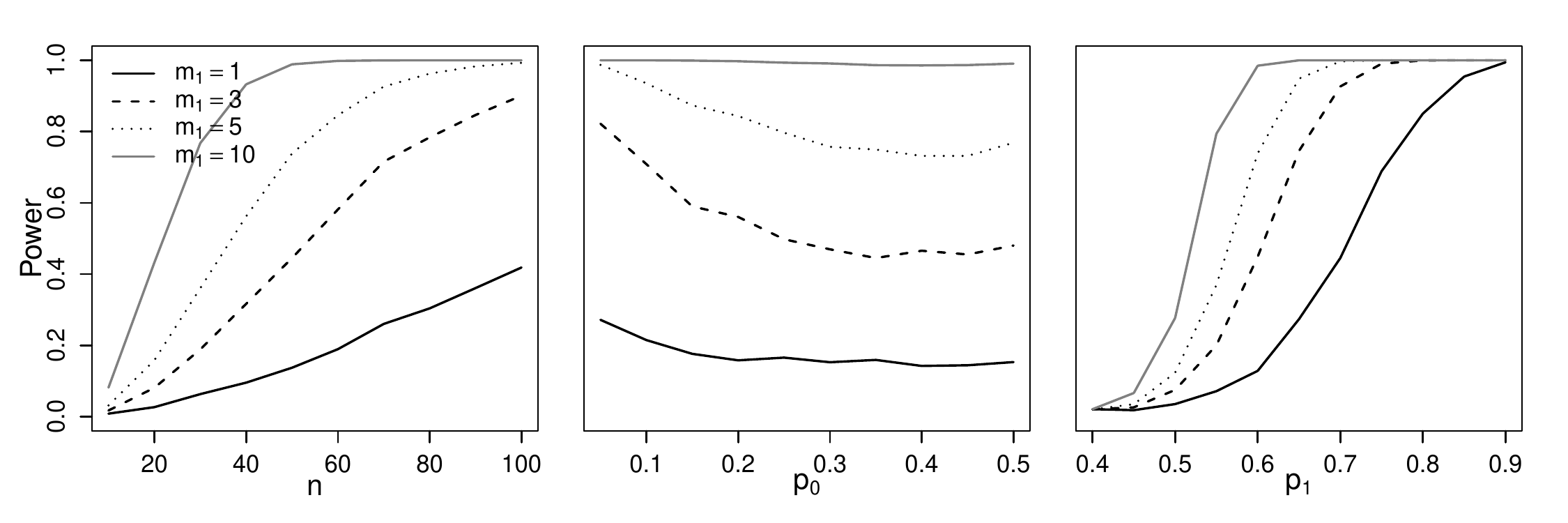}% %[height=10]
\caption{Power values for discrete data simulation for omnibus log p for $m_1=\{1,3,5,10\}$, $m=10$. The left chart shows results for increasing $n$ and constant $p_0=0.4$, $p_1=0.6$; the chart in the middle, constant $n=50$, increasing $p_0$ and $p_1=p_0+0.2$; The right chart $n=50$, $p_0=0.4$ and increasing $p_1$.}\label{fig:05}
\end{figure}

\section{Examples}

\subsection{Meta Analysis}

In meta analysis the evidence from several studies on a topic is combined. There are several examples in the literature, showing that the 
efficacy of a treatment can vary among studies. Reasons for such a variation can among other factors be differences in the underlying study populations, 
or environmental factors. If effect size estimates are available for all considered studies, 
a random effect meta analysis is often carried out. Global tests, such as the Fisher and the Stouffer test are 
a popular alternative option that does not require effect size estimates.

As an illustration, we applied our omnibus test to a data set from a meta analysis provided by the R-package metafor \citep{metafor}. We chose the data set dat.fine1993 where results from 17 studies are presented which compare post operative radiation therapy with or without adjuvant chemotherapy in patients with malignant gliomas \citep{Fine}. For each study the data set specifies the number of patients in the experimental group (receiving radiotherapy plus adjuvant chemotherapy) as well as the number of patients in the control group (receiving radiotherapy alone). 
In addition the number of survivors after 6, 12, 18, and 24 months follow-up within each group is given. 
One of the 17 studies recorded survival only at 12 and 24 months. For illustration purposes we performed a separate meta analysis for each time point and calculated a $\chi^2$-test (or Fisher's exact test, where appropriate) for each study. The resulting p-values were then applied to test the 
global null hypothesis using the following methods: Bonferroni, Simes, Fisher, Stouffer, higher critisism, and omnibus log p.

Table \ref{tab2a} shows the resulting p-values for the global tests. Note that the table does not display the results for Stouffer's method which in all cases results in a p-value close to 1 and will not be discussed further. As in the simulation study, the omnibus method is among the top methods for all time points except for the 12 months data, were the p-value of the Fisher combination test is approximately one third smaller than the p-values of the omnibus method. For the 6 months data, however, the advantage of the omnibus method as well as Bonferroni and Simes methods (all p-values between 0.12 and 0.13) 
over the Fisher test (p-value: 0.51) is considerable. The largest p-value across all scenarios turns out to be smallest for the omnibus test.

\ \begin{table}[h!]
\caption{\label{tab2a}
Meta analysis example I. Global tests have been applied to a meta analysis comparing post operative radiation therapy with or without adjuvant 
chemotherapy in patients with malignant gliomas. The p-values of the methods are shown when testing the global null hypothesis 
at different time points.
}
\begin{tabular}{c|ccccccc}
&omnibus&&& &\\
&  $log p$ &Bonferroni & Simes& Fisher &   HC\\ \hline
6 months&0.119  & 0.118 &0.118 & 0.509 &        0.500 \\
12 months& 0.257 &0.406 &0.235 & 0.178  &      0.355\\
18 months& 0.116 &0.279& 0.279 & 0.094     &   0.152\\
24 months&   0.013  &0.006 &0.006 & 0.019   &   0.033\\  \end{tabular}
\end{table}

We next analysed the data examples from the R-package metap \citep{metap}. We used five of the eight different data examples, ignoring three 
that involve only hypothetical data. For each of these data sets a vector of p-values of lengths ranging from 9 to 34 is provided in the package.
For instance the data taken from the meta analysis by \cite{Sutton} involves 34 randomized clinical trials where 
cholesterol lowering interventions were compared between treatment and control groups. 
The actual treatments were mostly drugs and diets. For each study, a test was performed to analyze, if the effect sizes (log Odds Ratio) 
are smaller than 0 (one-sided test), and p-values were calculated based on the normal distribution (\cite{Sutton}, Table 14.3). For details on the other data sets we refer to the original publications, for references see the documentation of the metap package. 
Note that for some studies p-values were derived from independent subgroup analyses. 

Table 3 compares again different tests of the global null in terms of their p-values. Three of the methods (Simes, Fisher, log p) 
lead to significant p-values at level $\alpha=0.05$ for four of the five data sets. The omnibus log p method however, is the only test that
also provides four significant results at level $\alpha=0.01$.

\ \begin{table}[h!]
\caption{\label{tab2}
Meta analysis example II. The table states the p-values obtained from several global null hypothesis tests. The underlying
data have been taken from the examples provided with the R-package {\em metap.}
}
\begin{tabular}{c|ccccccccc}
&omnibus&&&& &  \\
& $log p$  &Bonferroni & Simes& Fisher &Stouffer &   HC\\ \hline
%beckerp&  0.07 &0.08&   0.08  & 0.047 &  0.06 &    0.25\\
Sutton &  0.24  &0.13 &  0.13 &  0.79 &  1 &    0.57\\
%edgington 0.516&&1 &0.40& 0.396 &0.161 & 0.4\\
mourning&  0.007 &0.07 &  0.04 &  0.017   &0.11 &    0.013\\
naep&      $<$0.001 &$<$0.001     &   $<$0.001        &$<$0.001    &   $<$0.001  &           0.056\\
%rosenthal& 0.018 & 0.05 &  0.05 &  0.011 &  0.008&   &    0.12\\
teach&  0.0007  &0.019 &  0.019&   0.0014&   0.0077  &   0.24\\
validity&   $<$0.001  & $<$0.001  &  $<$0.0001&  $<$0.001  & $<$0.001  &     0.025\\
  \end{tabular}

\end{table}

\subsection{Experimental Evolution}

With the development of large scale inexpensive sequencing technologies, experiments became popular that aim to elucidate biological adaptation
at the molecular level of DNA and RNA. In such experiments, organisms are often exposed to stress factors for several generations, and their 
genetic adaptation is studied. With microorganisms, such stress factors can for instance result from antibiotics, with the adaptation being resistance.
With higher organisms examples of stress factors are temperature, or toxic substances.
 While evolution in nature usually takes place 
only once under comparable circumstances, experimental evolution can be done with replicate populations. Among other things,
replication permits to investigate the reproducibility of adaptation, a key topic in evolutionary genetics. 
The statistical challenge is to identify genomic positions (called loci) involved in adaptation. There is a large number of candidate loci, for which adaptation has to be distinguished from 
random temporal allele frequency changes due to 
genetic drift, as well as sampling and sequencing noise. 

Furthermore, recent research suggests that replicate populations often do not show a 
consistent behavior, with signals of adaptation showing up partially at different loci.
Two biological explanations for this finding is that beneficial alleles may be lost due to drift, and that 
the same adaptation at a phenotypic level
can often be achieved in multiple ways at the genomic level.

When testing for significant allele frequency changes, 
a test like our omnibus test is therefore desirable, as it enjoys good power also when
signals of adaptation are not consistent across replicates. 
We illustrate the application of our omnibus $log p$ test to data from an experiment on {\em Drosophila}
described in \citet{Griffin}.

\begin{figure}[!ht]%figureG1
\centering
\includegraphics[width=0.8\textwidth]{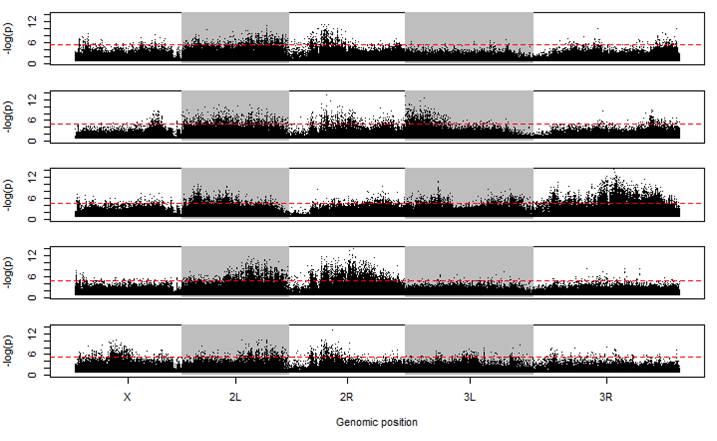}
\caption{Manhattan plots of negative logarithm of p-values from a genome wide scan of 
five replicate populations. The p-values have been corrected for multiple testing. Data are taken from
\citet{Griffin}.
}\label{figG01}
\end{figure}

\begin{figure}[!ht]%figureG2
\centering
\includegraphics[width=0.8\textwidth]{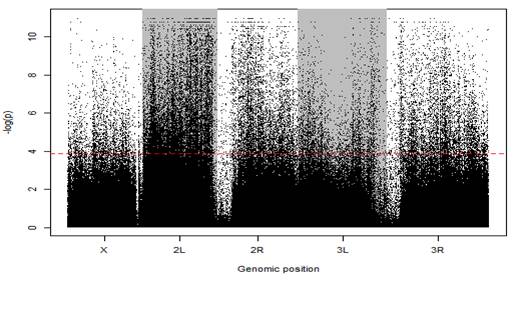}
\caption{Plot of combined evidence across replicates. Manhattan plots of the negative logarithm of the
p-values obtained with our (log-p) omnibus test. 
The p-values have been corrected for multiple testing. Data are taken from
\citet{Griffin}.
}\label{figG02}
\end{figure}

%CMH-test with correction, Metaanalysis, analysis of different subgroups

%\paragraph{Global test for binary variables: Cochrane-Mantel-Haenszel test (CMH)} Consider the case of two binary variables where the observations are further grouped in $s$ strata. The aim is to test if the two variables are independent within the $s$ strata, i.e., there is no consistent difference in proportion in the $s$ 2x2 tables. The null hypothesis states that the common odds ratio is equal to 1. The power of the CMH test is maximized, when the odds ratios are equal for all groups.

\section{Discussion}

In this manuscript we introduced new non-parametric omnibus tests for testing the global null hypothesis. Our proposed
approach enjoys very good power properties, no matter in how many cases the alternative holds.
In our comparison with alternative approaches, it is not always the best method,
but we did not find scenarios, where the omnibus test performs considerably worse than
the best alternative method for a given setup (as it is the case, e.g., for Bonferroni and Simes 
for large number of alternative hypotheses or Fisher and Stouffer for small number of alternative hypotheses).

For our test, we compute successive cumulative sums of the suitably transformed sorted individual p-values. 
The most unusual cumulative sum is then obtained by computing the p-value of each sum 
under the global null hypothesis. The smallest p-value is then used as test statistic.

We considered different transformations of the initial p-values $p_i$, in particular $1-p_i$, $-\log(p_i)$,
$\Phi^{-1}(1-p_i)$, and $p_i^{-1/2}$. Our results showed only small differences in power between the transformations.
However, the log p transfrom seems to lead to a particularly good trade off in power across many scenarios.

As expected the Simes test outperforms the Bonferroni procedure (or is equal) in the simulation study, though, for the considered scenarios the improvement in power is not remarkable. 

All our simulations are based on one-sided tests, but the methods also work for the two-sided testing scenario. For two-sided tests however, it is also possible to reject the global null hypothesis even when the individual hypotheses show clear effects in differing directions. 

%% BibTeX support
\bibliographystyle{plainnat}
\bibliography{bibliography}

\begin{thebibliography}{17}
\providecommand{\natexlab}[1]{#1}
\providecommand{\url}[1]{\texttt{#1}}
\expandafter\ifx\csname urlstyle\endcsname\relax
  \providecommand{\doi}[1]{doi: #1}\else
  \providecommand{\doi}{doi: \begingroup \urlstyle{rm}\Url}\fi

\bibitem[Crocetta and Loperfido(2005)]{Crocetta}
C.~Crocetta and N.~Loperfido.
\newblock The exact sampling of l-statistics.
\newblock \emph{METRON - International Journal of Statistics}, 63\penalty0
  (2):\penalty0 1--11, 2005.

\bibitem[Dickhaus(2014)]{Dickhaus}
T.~Dickhaus.
\newblock \emph{Simultaneous Statistical Inference. With applications in the
  Life Sciences}.
\newblock Heidelberg, Springer, 2014.

\bibitem[Donoho and Jin(2004)]{Don-Jin1}
D.~Donoho and J.~Jin.
\newblock Higher criticism for detecting sparse heterogeneous mixtures.
\newblock \emph{Annals of Statistics}, Jun 1:\penalty0 962--94, 2004.

\bibitem[Donoho and Jin(2015)]{Don-Jin2}
D.~Donoho and J.~Jin.
\newblock Higher criticism for large-scale inference, especially for rare and
  weak effects.
\newblock \emph{Statistical Science}, 30(1):\penalty0 1--25, 2015.

\bibitem[Fine et~al.(1993)Fine, Dear, Loeffler, Black, and Canellos]{Fine}
H.~A. Fine, K.~B. Dear, J.~S. Loeffler, P.~M. Black, and G.~P. Canellos.
\newblock Meta-analysis of radiation therapy with and without adjuvant
  chemotherapy for malignant gliomas in adults.
\newblock \emph{Cancer}, 71:\penalty0 2585--2597, 1993.

\bibitem[Fisher(1932)]{Fisher32}
R.~A. Fisher.
\newblock \emph{Statistical Methods for Research Workers}.
\newblock Oliver and Boyd, London, 1932.

\bibitem[Griffin et~al.(2017)Griffin, Hangartner, Fournier-Level, and
  Hoffmann]{Griffin}
Philippa~C. Griffin, Sandra~B. Hangartner, Alexandre Fournier-Level, and Ary~A.
  Hoffmann.
\newblock Genomic trajectories to desiccation resistance: Convergence and
  divergence among replicate selected drosophila lines.
\newblock \emph{Genetics}, 205\penalty0 (2):\penalty0 871--890, 2017.
\newblock ISSN 0016-6731.
\newblock \doi{10.1534/genetics.116.187104}.
\newblock URL \url{http://www.genetics.org/content/205/2/871}.

\bibitem[Haibe-Kains et~al.(2008)Haibe-Kains, Desmedt, Sotiriou, and
  Bontempi]{survcomp}
B.~Haibe-Kains, C.~Desmedt, C.~Sotiriou, and G.~Bontempi.
\newblock A comparative study of survival models for breast cancer
  prognostication based on microarray data: does a single gene beat them all?
\newblock \emph{Bioinformatics}, 24\penalty0 (19):\penalty0 2200--2208, 2008.

\bibitem[Ingster and Lepski(2003)]{ING}
Yu~Ingster and O~Lepski.
\newblock Multichannel nonparametric signal detection.
\newblock \emph{Mathematical Methods of Statistics}, 12\penalty0 (3):\penalty0
  247--275, 2003.

\bibitem[Nagaraja(2006)]{Nagaraja2006}
H.~N. Nagaraja.
\newblock \emph{Order Statistics from Independent Exponential Random Variables
  and the Sum of the Top Order Statistics}, pages 173--185.
\newblock Birkh{\"a}user Boston, Boston, MA, 2006.

\bibitem[Simes(1986)]{Simes86}
R.~J. Simes.
\newblock An improved bonferroni procedure for multiple tests of significance.
\newblock \emph{Biometrika}, 73\penalty0 (3):\penalty0 751--754, 1986.
\newblock ISSN 0006-3444.

\bibitem[Stigler(1969)]{Stigler69}
S.~M. Stigler.
\newblock Linear functions of order statistics.
\newblock \emph{Ann. Math. Statistics}, 40\penalty0 (3):\penalty0 770--788,
  1969.

\bibitem[Stouffer et~al.(1949)Stouffer, Suchman, DeVinney, Star, and
  Williams]{Stouffer}
S.~Stouffer, E.~Suchman, L.~DeVinney, S.~Star, and R.J. Williams.
\newblock The american soldier, vol.1: Adjustment during army life.
\newblock \emph{Princeton University Press, Princeton, USA}, 1949.

\bibitem[Sutton et~al.(2000)Sutton, Abrams, Jones, Sheldon, and Song]{Sutton}
A.J. Sutton, K.R. Abrams, D.R. Jones, T.A. Sheldon, and F.~Song.
\newblock \emph{Methods for Meta-Analysis in Medical Research}.
\newblock John Wiley a. Sons, Ltd, London, 2000.

\bibitem[Team(2015)]{R}
R~Core Team.
\newblock \emph{R: A Language and Environment for Statistical Computing}.
\newblock R Foundation for Statistical Computing, Vienna, Austria, 2015.
\newblock URL \url{http://www.R-project.org/}.

\bibitem[Tukey(1976)]{Tukey}
J.W. Tukey.
\newblock T13: N the higher criticism. course notes.
\newblock \emph{Statistics 411, Princeton University}, 1976.

\bibitem[Viechtbauer(2010)]{metafor}
W.~Viechtbauer.
\newblock Conducting meta-analyses in {R} with the {metafor} package.
\newblock \emph{Journal of Statistical Software}, 36\penalty0 (3):\penalty0
  1--48, 2010.
\newblock URL \url{http://www.jstatsoft.org/v36/i03/}.

\end{thebibliography}

\end{document}